# Defect-engineered scaling of lead-free ferroelectrics with ultralow-voltage switching

*Reza Ghanbari[1], Jiayue Wang[2,3], Harikrishnan KP[4], Zixiao Shi[5], Aarushi Khandelwal[2,3], Konnor Koons[1], Eli Rodrigues[1], Tao Zhou[6], Martin Holt[6], David A. Muller[4,7], Harold Y. Hwang[2,3], Ruijuan Xu[1]**

[1] Department of Materials Science and Engineering, North Carolina State University, Raleigh, NC, 27606, USA

[2] Department of Applied Physics, Stanford University, Stanford, CA, 94305, USA

[3] Stanford Institute for Materials and Energy Sciences, SLAC National Accelerator Laboratory, Menlo Park, CA, 94025, USA

[4] School of Applied and Engineering Physics, Cornell University, Ithaca, NY, 14853, USA

[5] Department of Chemistry and Chemical Biology, Cornell University, Ithaca, NY, 14850, USA

[6] Center for Nanoscale Materials, Argonne National Laboratory, Lemont, IL, 60517, USA

[7] Kavli Institute at Cornell for Nanoscale Science, Cornell University, Ithaca, NY, 14853, USA

*Email: rxu22@ncsu.edu

**Abstract:** Scaling ferroelectrics to nanometer thicknesses remains a central challenge for low-power, nonvolatile electronics, as leakage currents increasingly dominate with reduced dimensions. Alkali-based, lead-free ferroelectrics offer an environmentally sustainable alternative to lead-based systems, yet their scaling is severely limited by leakage arising from volatile alkali constituents. Here, we show that this intrinsic limitation can be transformed into an advantageous degree of freedom through defect engineering. By precisely modulating alkali deficiency during thin-film synthesis, we engineer clustered defect complexes that function as deep trap states, strongly suppressing leakage and enabling robust ferroelectric operation in ultrathin films down to the sub-10 nm regime at voltages below 100 mV. Our results establish defect-enabled scaling as a viable pathway for advancing environmentally benign ferroelectrics toward ultra-low-power, non-volatile electronic technologies.

## 1. Introduction

Continued scaling of microelectronic systems places increasingly stringent demands on dielectric and ferroelectric materials that operate reliably at nanometer dimensions while maintaining low switching energy and minimal leakage.[1-3] As the dimension of ferroelectrics approaches the ultrathin limit, electrostatic effects such as depolarization fields progressively destabilize switchable polarization.[4-6] Simultaneously, leakage currents rise sharply due to the amplified influence of defects and interfaces, undermining robust ferroelectric switching in ultrathin films.[7] Leakage has thus emerged as a universal bottleneck in ferroelectric scaling, constraining energy-efficient switching in logic devices and limiting capacitance density, breakdown strength, and long-term reliability in embedded thin-film capacitors.[8] Overcoming leakage while preserving robust switchable polarization in ultrathin ferroelectrics is essential for advancing low-power logic devices, nonvolatile memories, and high-density embedded capacitors for next-generation microelectronics.

In parallel with these scaling challenges, the push towards environmentally sustainable electronics has called for the search for alternatives to conventional lead-based ferroelectrics.[9, 10] Alkali-based, lead-free (anti-)ferroelectrics, including $LiNbO_3$, $KNbO_3$, $(K,Na)NbO_3$ (KNN), $NaNbO_3$, $KTaO_3$, and $(Bi,Na)TiO_3$, constitute a technologically important materials family that provides a versatile platform for next-generation electronic and photonic devices. These oxides exhibit a broad spectrum of ferroelectric, dielectric, piezoelectric, and electro-optic functionalities, combined with rich structural phases and properties that are broadly tunable through controlling growth conditions, composition, temperature, strain, and other external stimuli.[11-23] Despite extensive investigation in bulk ceramics and single crystals, these materials remain comparatively underexplored in thin-film form, which is essential for modern microelectronic integration. In particular, the high volatility of alkali species in these materials leads to pronounced non-stoichiometry and defect formation,[24, 25] making leakage control increasingly difficult particularly as film thickness is reduced. While defect-engineering strategies have proven effective for controlling polarization direction, enhancing ferroelectric transition temperatures, and suppressing leakage in ferroelectric thin films,[26-28] their application to alkali-based lead-free systems remains limited. Moreover, commonly used defect engineering approaches often involve intrinsic trade-offs: defects introduced during synthesis or via ion bombardment act as pinning sites that hinder domain wall motion, leading to increased coercive fields and imprint,[28-30] whereas defects introduced via chemical doping increase compositional complexity and structural disorder.[31, 32] Developing scalable strategies to suppress leakage in ultrathin, lead-free, alkali-based ferroelectrics without compromising ferroelectric switching therefore remains an open challenge.

Here, we show that the intrinsic volatility of alkali species, long regarded as a fundamental limitation, can instead be leveraged as an advantageous degree of freedom for defect-controlled scaling of lead-free ferroelectrics. By precisely modulating alkali stoichiometry *in situ* during thin-film synthesis, we induce the formation of clustered defect complexes that act as deep trap states, strongly suppressing leakage while preserving robust, imprint-free ferroelectric switching. This approach enables reliable ferroelectric operation down to the sub-10 nm regime with ultralow voltage switching below 100 mV. When combined with strain engineering through growth on highly compressive substrates, the remanent polarization is further enhanced up to 26%, compared with low-strain-state films. Our results establish defect-enabled scaling as a viable pathway for advancing environmentally benign ferroelectrics toward ultra-low-power, non-volatile electronic technologies.

## 2. Results and Discussion

### 2.1. In situ control of leakage currents

Controlling alkali stoichiometry and defect formation requires a growth parameter capable of directly regulating alkali volatility during synthesis. In pulsed laser deposition (PLD), laser fluence governs ablation chemistry and growth kinetics, offering a possible route to tune alkali stoichiometry. To begin, we investigate how laser fluence influences the structural evolution of alkali-based ferroelectric thin films, using $NaNbO_3$ (NNO) as a model system. We synthesize epitaxial NNO films with a thickness of 80 nm on (001)-oriented $SrTiO_3$ (STO) single-crystal substrates while systematically varying laser fluences from 1.3 to 2.1 J $cm^{-2}$. Regardless of fluence, all films exhibit high crystalline quality, as evidenced by the presence of pronounced Laue fringes in high-resolution X-ray $\theta$-$2\theta$ scans (Figure 1a). To assess whether laser fluence influences the phase structures of NNO, we perform synchrotron-based 3D reciprocal space mapping (3D-RSM). Our previous work[13] demonstrates that epitaxial NNO films grown on STO can exhibit a strain-induced morphotropic phase boundary-like mixed-phase structure, consisting of coexisting monoclinic $M_B$ (*Cc*) and $M_C$ (*Cm* or *Pc*) phases which persists even at reduced thicknesses. Consistent with this, all films, independent of fluence, exhibit mixed-phase characteristics, with the 002 diffraction of $M_B$ and $M_C$ located at higher and lower L values relative to STO, respectively, indicating that laser fluence has minimal impact on the phase structure (Figure. 1b and Figure S1). This is further corroborated by piezoresponse force microscopy (PFM) imaging, which reveals domain patterns characteristic of $M_B$ - $M_C$ phases coexistence in all films, with only slight variations in their relative phase fractions (Figure S2). PFM topography confirms atomically smooth surfaces with root-mean-square (RMS) roughness values ranging from 186 to 251 pm (Figure S2). Yet, a more detailed quantitative analysis of the RSM data reveals a systematic expansion of the lattice parameter associated with the $M_C$ phase as laser fluence increases, whereas the $M_B$ phase shows a less pronounced dependence. By combining the phase fractions extracted from PFM with the corresponding lattice parameters, we estimate an effective average lattice parameter, which exhibits a slight overall lattice expansion with increasing fluence (Figure 1c).

To elucidate the effect of laser fluence on electrical behavior, we measure polarization-electric field (*P-E*) hysteresis loops and leakage currents using a vertical capacitor geometry with symmetric epitaxial $La_{0.7}Sr_{0.3}MnO_3$ electrodes. Surprisingly, despite only subtle fluence-dependent structural variations, the electrical properties exhibit a strong dependence on laser fluence. In particular, the leakage current decreases dramatically, by 5 orders of magnitude from 10 to $10^{-4}$ A $cm^{-2}$ at an electric field of 100 kV $cm^{-1}$, as fluence increases (Figure 1d). This pronounced leakage suppression drives a transformation from leakage-dominated, unsaturated ferroelectric hysteresis loops at low fluence conditions to well-saturated, symmetric ferroelectric switching at high fluence conditions (Figure 1e). These results demonstrate that laser fluence plays a dominant role in governing the electrical performance of NNO films, which is independent of the mixed-phase structure in NNO.

### 2.2. Defect-mediated conduction mechanisms

What gives rise to the pronounced fluence-dependent differences in leakage behavior? To address this question, we first investigate the conduction mechanisms in low- and high-fluence NNO films using temperature-dependent current-voltage (I-V) measurements (Figure 2, Figure S3 and Supplementary Note 1). This approach is widely employed to identify dominant trap states and to distinguish among common conduction mechanisms (e.g., Ohmic, Schottky, Poole-Frenkel, modified Poole-Frenkel emission, and space charge limited conduction) in dielectric oxides.[28, 29, 33-35] Our analysis focuses on two representative samples: a low-fluence film grown at 1.5 J $cm^{-2}$ with substantially higher leakage (Figure 2a), and a high-fluence film grown at 2.1 J $cm^{-2}$, which exhibits the lowest leakage current (Figure 2b). Note that the sample grown at the lowest fluence (1.3 J $cm^{-2}$) was excluded from this analysis due to excessive leakage at elevated temperatures, which precludes reliable fitting. For the low-fluence NNO film, the I-V characteristics are well fitted by Poole-Frenkel emission, a trap-assisted transport mechanism governed by the thermal emission of charge carriers from shallow trap states, as evidenced by a scaling factor ($r$) of approximately 1 (Figure 2c). In contrast, the high-fluence NNO film exhibits more complex transport behavior: while its low-field conduction can be well fitted by standard Poole-Frenkel emission, at electric fields exceeding ~250 kV $cm^{-1}$, the conduction deviates from conventional Poole-Frenkel behavior, with a scaling factor ranging between 1 and 2 (Figure 2d). This behavior is characteristic of modified Poole-Frenkel emission, which is typically associated with a complex, spatially distributed deeper trap states rather than isolated shallow traps. To further quantify the trap energy, we extract activation energies from Arrhenius plots of leakage current density (σ) as a function of inverse temperature (1000/T) at fixed applied voltages, followed by extrapolation to obtain the zero-field ionization energy (Figure 2e, f). For the low-fluence NNO film, the extracted activation energies ($E_V$) fall within the range of 0.53–0.59 eV, yielding a zero-field ionization energy ($E_I$) of approximately 0.53 eV (Figure 2e). By contrast, the high-fluence NNO film exhibits two distinct regimes: at low fields, activation energies ($E_V$) of 0.53–0.55 eV and a zero-field ionization energy ($E_I$) of 0.53 eV, consistent with the standard Poole-Frenkel mechanism; and at high fields, significantly higher activation energies of 0.78–0.84 eV and a zero-field ionization energy ($E_I$) of 0.845 eV, indicative of deeper trap states associated with modified Poole-Frenkel conduction (Figure 2f). Thickness-dependent fitting of high-fluence films further confirms that modified Poole-Frenkel emission remains the dominant conduction mechanism across the measured thickness range (Figure S4).

To understand the microscopic origin of the distinct trap states observed in low- and high-fluence NNO films, we compare the extrapolated zero-field ionization energies with prior theoretical predictions.[36] In low-fluence films, the dominant shallow trap states associated with Poole-Frenkel emission exhibit ionization energies close to the calculated activation energy for isolated Na vacancies (~0.6 eV). These isolated vacancy defects act as shallow acceptor-type defects, providing electrostatically favorable sites for hole trapping in niobate perovskites. Such shallow trap states can promote thermally activated hole conduction, leading to increased leakage currents and degraded ferroelectric switching behavior in low fluence NNO films (Figure 1d, e). By contrast, the significantly higher ionization energies measured in high-fluence films indicate a transition from shallow to deep trap states. These deep traps might be related to the clustering of Nb antisite-based defect complexes, which are known to form in niobate perovskites under alkali-deficient conditions due to excess Nb incorporation.[20, 37] The introduction of these deep-lying trap states strongly localizes charge carriers, effectively suppressing leakage currents and enabling robust ferroelectric switching in high-fluence NNO films. In comparison, isolated Na vacancies and vacancy-only complexes, such as $V_{Na}$-$V_O$-$V_{Na}$, may still coexist in the films but are predicted to introduce relatively shallow trap states,[36] making them unlikely to be the primary origin of the deeper trap states and suppressed leakage observed in high-fluence films. Consistent with this interpretation, oxygen annealing of 80 nm NNO heterostructures at 600 °C for 2 h in 760 Torr $O_2$ produced no obvious change in the leakage current or P-E hysteresis loops (Figure S5), suggesting that isolated oxygen vacancies are unlikely to be the primary origin of the deep traps observed in the high-fluence films.

## 2.3. Atomic-scale understanding of defect structures

To further elucidate the defect nature of these trap states and their correlation with laser fluence, we analyze the elemental composition and stoichiometry as a function of laser fluence using X-ray photoelectron spectroscopy (XPS) (Figure 3a and Figure S6). The XPS results reveal a systematic decrease in the Na/Nb ratio with increasing laser fluence, indicating a transition from Na-rich to stoichiometric and ultimately to Na-deficient compositions, reaching approximately 33% Na deficiency at 2.1 J $cm^{-2}$. Several mechanisms are likely to contribute to this trend: first, higher laser fluence increases the kinetic energy of ablated species, enhancing knock-on damage during growth and promoting Na cation vacancy formation, given the relatively low formation energy of Na vacancies in NNO.[38] Second, higher laser fluence increases the local temperature of the ablation plume, promoting preferential Na evaporation and generating a Na-deficient plume that transfers its non-stoichiometry to the growing film. Additional XPS measurements on high-fluence films with different thicknesses (25-185 nm) show only slight variation in Na content (Figure S7), supporting reproducible fluence-controlled stoichiometry across the thickness series. Under such highly alkali-deficient growth conditions, excessive Nb results in a negative formation energy of Nb antisites,[37] making it energetically favorable for Nb to occupy Na vacancy sites.

Incorporation of Nb on the A-site modifies the local bonding environment and leads to a partial reduction of Nb from its nominal +5 oxidation state. This behavior is directly confirmed by XPS measurements, which show a systematic shift of the Nb 3d peaks toward lower binding energies with increasing fluence (Figure 3b). The Nb antisites carry an effective charge of +4; to maintain charge neutrality, this defect must be associated with negatively charged A-site defects, either through coupling with four Na vacancies or by forming an antisite pair with a Na atom occupying the B-site ($Na_{Nb}$).[39] However, the formation of Na antisites is highly unlikely under the severe Na-deficient conditions present in high-fluence films. As a result, the lattice preferentially stabilizes $Nb_{Na}^{\cdot\cdot\cdot\cdot} - 4V_{Na}'$ defect complexes as the dominant defect configuration.

To confirm the presence of these defect complexes, we conduct scanning transmission electron microscopy (STEM) on representative samples: low- and high-fluence NNO films grown at 1.3 J cm$^{-2}$ and 2.1 J cm$^{-2}$, respectively. In low-fluence films, cross-sectional low-angle annular dark-field (LAADF)-STEM images exhibit uniform contrast across the film thickness (Figure 3c), indicative of a structurally homogeneous lattice with minimal defect density. In contrast, high-fluence films reveal the emergence of vertically aligned columnar features embedded within the perovskite matrix (Figure 3d), suggesting local deviations from ideal cation ordering due to the clustering of defect complexes. Atomically-resolved high-angle annular dark-field (HAADF) imaging further reveals Nb atoms occupying nominal Na sites (Figure 3e), leading to the formation of additional Nb columns within locally distorted regions, as illustrated schematically. Similar defect clustering has been predicted in other alkali-deficient niobate systems.[39, 40] Corresponding strain maps reveal pronounced strain concentration around these defect complexes (Figure 3f and Figure S8), arising from the substantial off-centering of Nb atoms and consistent with the slight lattice expansion observed in high-fluence films (Figure 1c). Taken together, these results establish a clear evolution of defect states with laser fluence: from isolated Na vacancies acting as shallow trap states at low fluence to $Nb_{Na}^{\cdot\cdot\cdot\cdot} - 4V_{Na}'$ defect complexes acting as deep trap states at high fluence, as summarized schematically (Figure 3g). The formation and clustering of the defect complexes generate pronounced local lattice distortions and deep electrostatic potential minimum, driving trap levels deeper into the bandgap. This defect landscape promotes strong carrier localization and substantially reduces carrier mobility, thereby effectively suppressing leakage currents.

## 2.4. Defect-engineered scaling of lead-free ferroelectrics

This fluence-controlled introduction of defect complexes effectively suppresses the leakage currents in lead-free NNO films. Leveraging this strategy, we systematically investigate the thickness scaling of ferroelectric switching and demonstrate robust ferroelectric switching with low leakage down to a thickness of 9 nm, enabling ultralow coercive voltages below 100 mV. Using high-fluence growth conditions, we synthesize NNO films with thicknesses ranging from 9 - 185 nm. P-E loop measurements reveal well-saturated and symmetric hysteresis loops over a broad frequency range from 1 Hz to 10 kHz across the entire thickness series (Figure 4a,b and Figure S9), indicating robust ferroelectric switching with minimal leakage. Positive-up negative-down (PUND) measurements in the ultrathin limit (9 and 18 nm capacitors) further confirm genuine switchable remanent polarization by separating switching responses from non-switching capacitive and leakage contributions (Figure S10). To further evaluate the electrical endurance of the films, we perform cyclic switching measurements using bipolar pulses. The switched polarization remains largely stable after more than $10^8$ switching cycles, with minimal fatigue observed across the investigated thickness range (Figure S11). Together, these frequency-dependent hysteresis and cyclic endurance measurements demonstrate the robust and stable ferroelectric response of the $NaNbO_3$ films. This marked improvement in thickness scalability directly originates from the effective suppression of leakage currents enabled by defect engineering (Figure S12). Notably, in contrast to ion bombardment-induced or as-grown aligned charged point defect complexes in ferroelectric thin films previously reported in ferroelectric thin films, which have often been associated with pronounced imprint behavior due to localized charged defects near the film-electrode interface or highly asymmetric defect distributions,[26-29] the clustered charge-neutral defect complexes formed in our films may be distributed more uniformly throughout the film thickness, resulting in the relatively small imprint observed in the P-E loops. It is important to note that the preferred out-of-plane polarization response observed in the vertical PFM phase images (Figure S2) should not be directly interpreted as defect-induced imprint but may instead be influenced by asymmetric electrostatic boundary condition associated with the presence of the LSMO bottom electrode in the as-grown heterostructures. Consistent with this interpretation, the nearly symmetric P-E hysteresis loops measured in capacitors with symmetric LSMO electrodes suggest the absence of a strong macroscopic internal bias field.

As the film thickness decreases, the remanent polarization gradually reduces due to the increasing influence of depolarization fields, causing the change in polarization from 42 μC cm$^{-2}$ in 80 nm films to 6 μC cm$^{-2}$ in 9 nm films (Figure 4c), However, possible secondary contributions such as interfacial dead layers cannot be fully excluded and require further atomic-scale structural and chemical characterization. Importantly, this reduction in remanent polarization is accompanied by a pronounced decrease in coercive field for films thinner than ~ 25 nm (Figure 4d). The observed thickness dependence of the coercive field deviates significantly from the classical Janovec–Kay–Dunn (JKD) scaling law ($E_c \propto t^{-2/3}$), which we attribute to the mixed-phase structure of MPB-NNO. The coexistence of monoclinic and triclinic phases facilitates polarization rotation and lowers the switching energy barrier, giving rise to sub-JKD scaling behavior. As a result, the reduced coercive field at small thicknesses enables sub-100 mV switching voltages, reaching values as low as 20 mV at a thickness of 9 nm (Figure 4c). Despite the reduced thickness, the remanent polarization of these ultrathin films remains within the practically sufficient range of 5-10 μC cm$^{-2}$ for device-relevant applications.[6]

Moreover, we combine this defect engineering strategy with strain engineering by synthesizing NNO films on highly compressive substrates to further enhance the achievable polarization. In addition to growth on STO, which imposes a small compressive strain of ~ -0.25%, we synthesize NNO films under high-fluence conditions on $La_{0.3}Sr_{0.7}Al_{0.65}Ta_{0.35}O_3$ (LSAT; a ≈ 3.868 Å) and $LaAlO_3$ (LAO; a ≈ 3.79 Å). Assuming fully strained films and using the pseudocubic lattice parameter of bulk NNO in its antiferroelectric ground state (a ≈ 3.88 Å), these substrates generate theoretical compressive strains of ~ -1.21 % and ~ -3.19 %, respectively. X-ray $\theta$–$2\theta$ scans reveal that NNO films grown on LSAT remain nearly fully strained from 25 nm up to 80 nm. In contrast, films grown on LAO experience more rapid strain relaxation at the larger lattice mismatch, precluding coherent strain at higher thicknesses. Nevertheless, 25 nm NNO films on LAO retain a substantial compressive strain of ~ -1.7%, corresponding to a lattice parameter of 3.968 Å for NNO films (Figure S13). These highly strained films exhibit enhanced remanent and saturated polarizations, as evidenced by well-saturated P-E loops with minimal leakage currents (Figures S14 and S15). Although RSM measurements for films grown on these substrates (Figure S7B) show no signature of $M_B$–$M_C$ phase coexistence (unlike films grown on STO), we perform a detailed study of conduction mechanisms in NNO films synthesized on LSAT (Figure S16a, b). The results confirm that the modified Poole–Frenkel conduction mechanism remains in these highly strained films, giving rise to an activation energy of 0.75–0.81 eV, consistent with those extracted for NNO films grown on STO (0.78–0.84 eV; Figure S16c). By combining enhanced compressive strain with effective leakage suppression via defect engineering, we achieve substantial enhancements in saturated polarization of 17% and 26% for films grown on LSAT and LAO, respectively, compared with films grown on STO (Figure 4e).

A low coercive field is essential for low-power logic applications that require low operating voltages, whereas a sufficiently high remanent polarization is critical for non-volatile memories that require robust polarization states. Plotting the coercive field as a function of remanent polarization therefore provides a direct metric to evaluate how closely a given material approaches the optimal performance window required for low-voltage, non-volatile ferroelectric devices. In Figure 4f, we benchmark the measured remanent polarization and coercive field of our NNO films against previously reported high-performance (001)-oriented perovskite ferroelectric thin films,[6, 41-53] including lead-based $PbZr_{1-x}Ti_xO_3$ (PZT) and representative lead-free systems such as $BaTiO_3$, $BiFeO_3$, and non-MPB NNO films. By combining leakage suppression with strain engineering, thickness-scaled NNO films span both ends of this performance spectrum. Thicker films exhibit robust remanent polarizations of ~ 30–50 μC $cm^{-2}$, comparable to or exceeding those of other lead-free ferroelectric thin films. While PZT offers higher remanent polarization, its lead content raises persistent environmental and health concerns, and its strong polarization state typically limits further reduction of coercive voltage. In contrast, as NNO film thickness is reduced, we achieve ultra-low switching voltage down to sub-100 mV while maintaining minimal leakage in the sub-10 nm thickness regime, comparable to or surpasses that of most reported lead-based and lead-free ferroelectric thin films. Together, these results demonstrate that defect-engineered NNO films uniquely combine a low coercive voltage with sufficiently high remanent polarization across a wide thickness range, positioning alkali-based lead-free ferroelectrics as a strong candidate for next-generation ferroelectric applications.

## 3. Conclusion

In summary, we demonstrate a defect-engineering strategy that suppresses leakage currents in ultrathin NNO films by transforming the alkali volatility from a fundamental limitation into a controllable materials parameter. Distinct from conventional defect-engineering approaches developed for lead-based ferroelectrics, this strategy enables robust, imprint-free ferroelectric switching at thicknesses below 10 nm and supports ultra-low-voltage operation down to sub-100 mV. Importantly, this approach is not specific to NNO but is broadly applicable to a wide class of technologically important alkali-based oxides, spanning dielectrics, ferroelectrics, electro-optics, including $KNbO_3$, KNN, $LiNbO_3$, $LiTaO_3$, and $KTaO_3$, etc. Although lead-free, alkali-based ferroelectric thin films remain underexplored compared to other lead-free systems such as $BaTiO_3$ and $BiFeO_3$, our results establish a viable materials pathway and provide a strong foundation for integrating these environmentally benign ferroelectrics into next-generation low-power, non-volatile microelectronic technologies. By enabling non-leaky ferroelectricity in the ultrathin limit within a lead-free materials platform, this defect-engineering approach can be utilized to address a key materials bottleneck for continued device-level scaling toward highly energy-efficient microelectronic devices.

## 4. Materials and Methods

### Thin-film synthesis

Epitaxial heterostructures were synthesized on (001)-oriented single-crystal STO substrates using pulsed laser deposition. First, a 10 nm-thick LSMO bottom electrode was deposited at a heater temperature of 730 °C, a dynamic oxygen pressure of 190 mTorr, a laser fluence of 1.6 J $cm^{-2}$, and a repetition rate of 3 Hz, using an imaged laser spot of 3.7 $mm^2$. Subsequently, NNO films with a thickness of 80 nm were synthesized in a dynamic oxygen pressure of 210 mTorr, at a heater temperature of 660 °C, and a repetition rate of 2 Hz, using an imaged laser spot of 4.58 $mm^2$, while systematically varying the laser fluence from 1.3 J $cm^{-2}$ to 2.1 J $cm^{-2}$. Thickness-dependent NNO films (9 -185 nm) were synthesized at a fixed laser fluence of 2.1 J $cm^{-2}$, while all other growth parameters held constant. For fabricating NNO capacitors, a 60 nm-thick LSMO top electrode layer was deposited following the growth of NNO, forming a tri-layer heterostructure. The top electrode layer was synthesized in a dynamic oxygen pressure of 190 mTorr, at a heater temperature of 660 °C, a laser fluence of 1.6 J $cm^{-2}$, and a repetition rate of 3 Hz, using an imaged laser spot of 3.7 $mm^2$. The heterostructures were subsequently cooled to room temperature at a rate of 5 °C $min^{-1}$ in a static oxygen pressure of 2.5 Torr.

### X-ray diffraction

$\theta$–$2\theta$ line scans were conducted using a high-resolution X-ray diffractometer (Rigaku SmartLab). Reciprocal Space Mapping was performed at the Center for Nanoscale Materials Hard X-ray Nanoprobe Beamline at the Advanced Photon Source. The beam energy was 10.5 keV. The footprint of the incident beam on the sample was 0.5 × 0.5 mm. The 3D reciprocal space around the (002) $NaNbO_3$ peak was measured by rocking the sample at a 0.01-degree step size while collecting 2D diffraction patterns on an Eiger 2 X 1M detector positioned 0.5 m from the sample. The RSM was reconstructed using a customized Python script, after removing the strong STO (002) peak in the vicinity.

### Piezoresponse force microscopy

The domain structure was imaged using Dual AC Resonance Tracking (DART) piezoresponse force microscopy. These measurements were performed with an MFP-3D Origin+ AFM (Asylum Research) with a conductive Pt/Ir-coated tip (Nanosensor, PPP-EFM, force constant ≈ 2.8 N $m^{-1}$).

**Electrical measurements**

Capacitor structures with symmetric LSMO top and bottom electrodes were fabricated using photolithography and chemical etching. Circular top electrodes with diameters of 10, 20, 50, and 75 μm were defined by patterning a photoresist layer on the as-grown heterostructures, followed by a selective removal of the exposed LSMO top electrode by immersing the heterostructures in a diluted phosphoric acid solution ($H_3PO_4$: DI water = 1:5) at 70 °C for 15 s. Following etching, the photoresist was dissolved by soaking the capacitor structures in acetone at 70 °C. Ferroelectric properties were characterized using a Precision Multiferroic Tester (Radiant Technologies). Temperature-dependent current–voltage measurements were performed using a thermal controller (Signatone S1080) coupled with the Radiant tester over a range of 292-420 K in 20 K increments and a ramp rate of 3 K $min^{-1}$, in both positive and negative polarities using an unswitched triangular voltage profile. The samples were stabilized for 15 min at each temperature before measurements. The reported values are averaged over eight capacitors to ensure reproducibility.

**X-ray photoelectron spectroscopy**

X-ray photoelectron spectroscopy (XPS) analyses were conducted using a PHI VersaProbe 4 equipped with a monochromatized Al Kα source. Spectra calibration was performed by aligning the adventitious carbon peak at 284.8 eV. The elemental stoichiometric ratios were quantified using the MultiPak software.

**Scanning Transmission Electron Microscopy**

The STEM lamellae were prepared using the standard lift-out method on a Thermo Fisher Helios G4 UX focused ion beam. The STEM imaging was performed on aberration-corrected Thermo Fisher Spectra 300 X-CFEG STEM devices operated at 300 kV, with a semi-convergence angle ($\alpha$) of 30 mrad. The collection angles for the LAADF and HAADF images are 30-55 mrad ($1\alpha$ - $1.8\alpha$) and 62 – 200 mrad ($2\alpha$ - $6.6\alpha$) respectively. The strain map is calculated from the HAADF image using the phase lock-in analysis on the primary FFT spot in the in-plane direction.[52, 53]

**Data availability statement**

All data supporting the findings of this study are available within the Article and its Supplementary Information.

**Supporting Information**

Supplementary Note 1 and Supplementary Figures 1-16.

**Acknowledgments**

R.G. and R.X. acknowledge support from the National Science Foundation (NSF) under award No. DMR-2442399 and the American Chemical Society Petroleum Research Fund under award No. 68244-DNI10. K. K. acknowledges support from the Army Research Office under award No. W911NF-25-1-0201. E. R. acknowledges support from the NC State College of Engineering (COE) Research Experience for Undergraduate (REU) program, funded by the COE Enhancement Fee. Work performed at the Center for Nanoscale Materials and Advanced Photon Source on APS beam time award(s) (DOI: https://doi.org/10.46936/APS-189892/60014012), both U.S. Department of Energy Office of Science User Facilities, was supported by the U.S. DOE, Office of Basic Energy Sciences, under Contract No. DE-AC02-06CH11357. The electron microscopy work made use of the Cornell Center for Materials Research shared instrumentation facility and FIB, STEM acquisition supported by the NSF (DMR-2039380). H.K. and D.A.M. acknowledge funding support from the Department of Defense, Army Research Office (ARO) under the ETHOS MURI award W911NF-21-2-0162. J.W., A.K., and H.Y.H. acknowledge support from the U.S. Department of Energy, Office of Basic Energy Sciences, Division of Materials Sciences and Engineering (DE-AC02-76SF00515).

## Main Figures

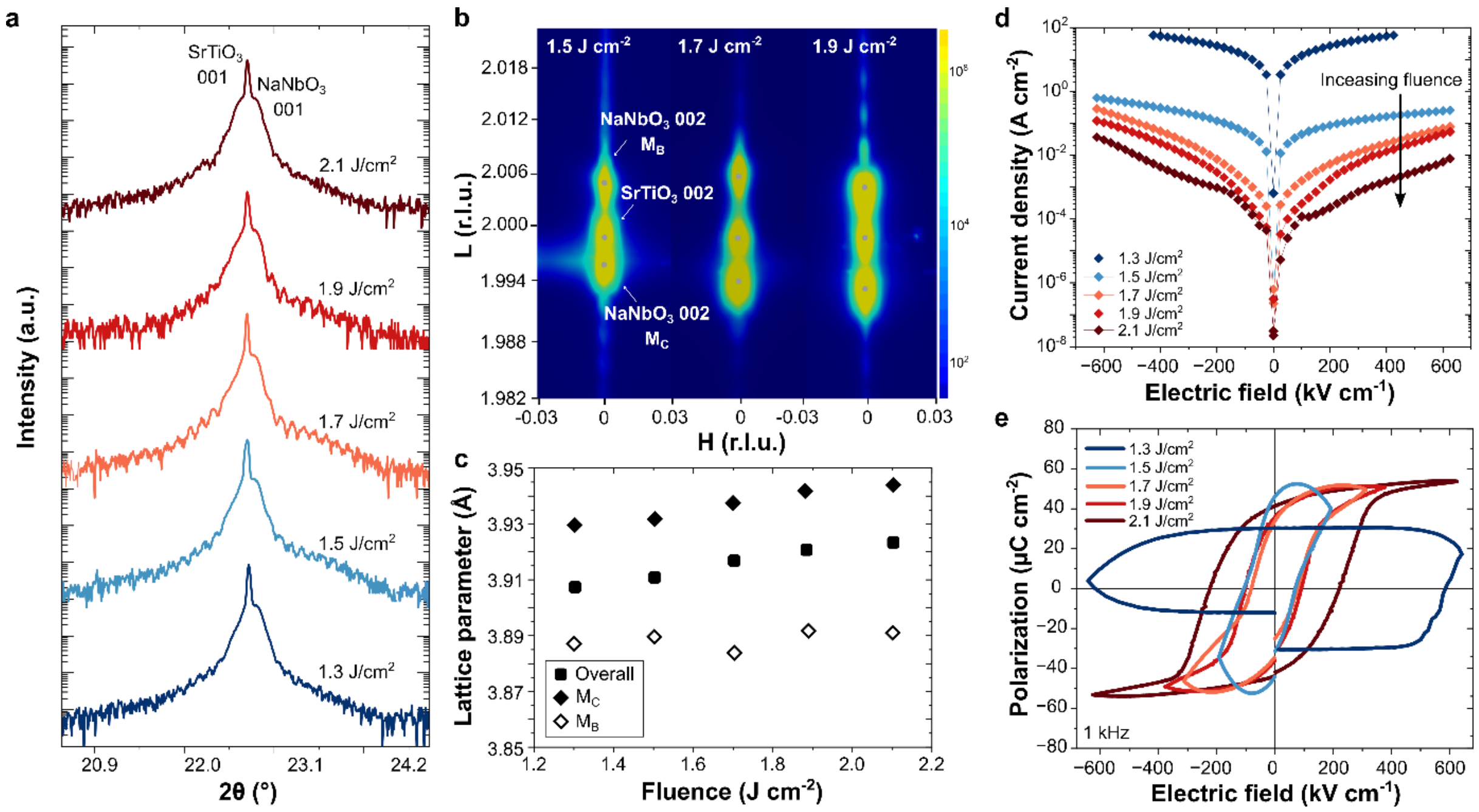

**Figure 1. Evolution of structure and ferroelectric properties in $NaNbO_3/SrTiO_3$ (001) heterostructures with laser fluence. (a)** X-ray $\theta$-$2\theta$ diffraction patterns near the pseudocubic 001-diffraction and **(b)** reciprocal space maps (RSMs) near the 002-diffraction for 80 nm thick $NaNbO_3/SrTiO_3$ (001) heterostructures as a function of laser fluence, confirming the phase coexistence throughout the entire studied fluence range. **(c)** Extracted out-of-plane *c*-lattice parameters for $M_B$ and $M_C$ phases, along with the averaged *c*-lattice parameter, as a function of fluence, revealing a slight lattice expansion with increasing fluence. **(d)** Leakage current-electric field measurements as a function of fluence, showing a pronounced reduction in leakage current with increasing fluence. **(e)** Polarization-electric field (*P*-*E*) hysteresis loops measured at 1 kHz, exhibiting increasingly well-saturated and symmetric ferroelectric switching with increasing fluence.

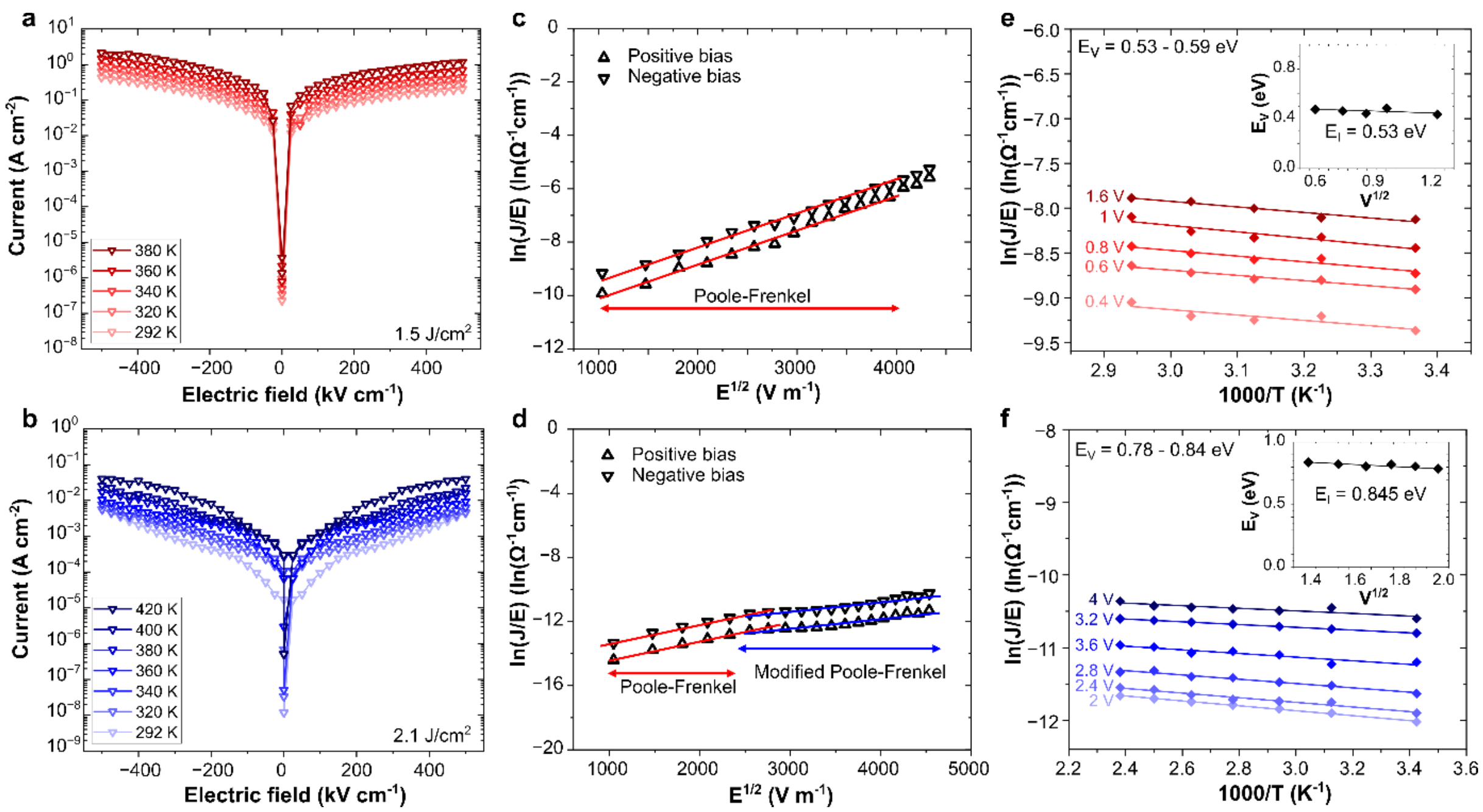

**Figure 2. Elucidating conduction mechanisms in $NaNbO_3$ films as a function of laser fluence.** Temperature-dependent leakage current-electric field measurements for $NaNbO_3$ films synthesized at **(a)** 1.5 J cm$^{-2}$ and **(b)** 2.1 J cm$^{-2}$ fluences. The leakage behavior is well described by **(c)** Poole-Frenkel emission ($r$ = 1, red solid lines) for films synthesized at 1.5 J cm$^{-2}$, and by **(d)** Poole-Frenkel emission in the low-field regime together with Modified Poole-Frenkel emission ($r$ = 1.94, blue solid lines) in the high-field regime for films synthesized at 2.1 J cm$^{-2}$. Arrhenius plots of ln (J/E) versus 1000/T at different applied voltages are shown for films synthesized at **(e)** 1.5 J cm$^{-2}$ and **(f)** 2.1 J cm$^{-2}$, from which the activation energies ($E_V$) are extracted from the slopes of the linear fits. Insets show the extrapolated zero-field trap ionization energies ($E_I$).

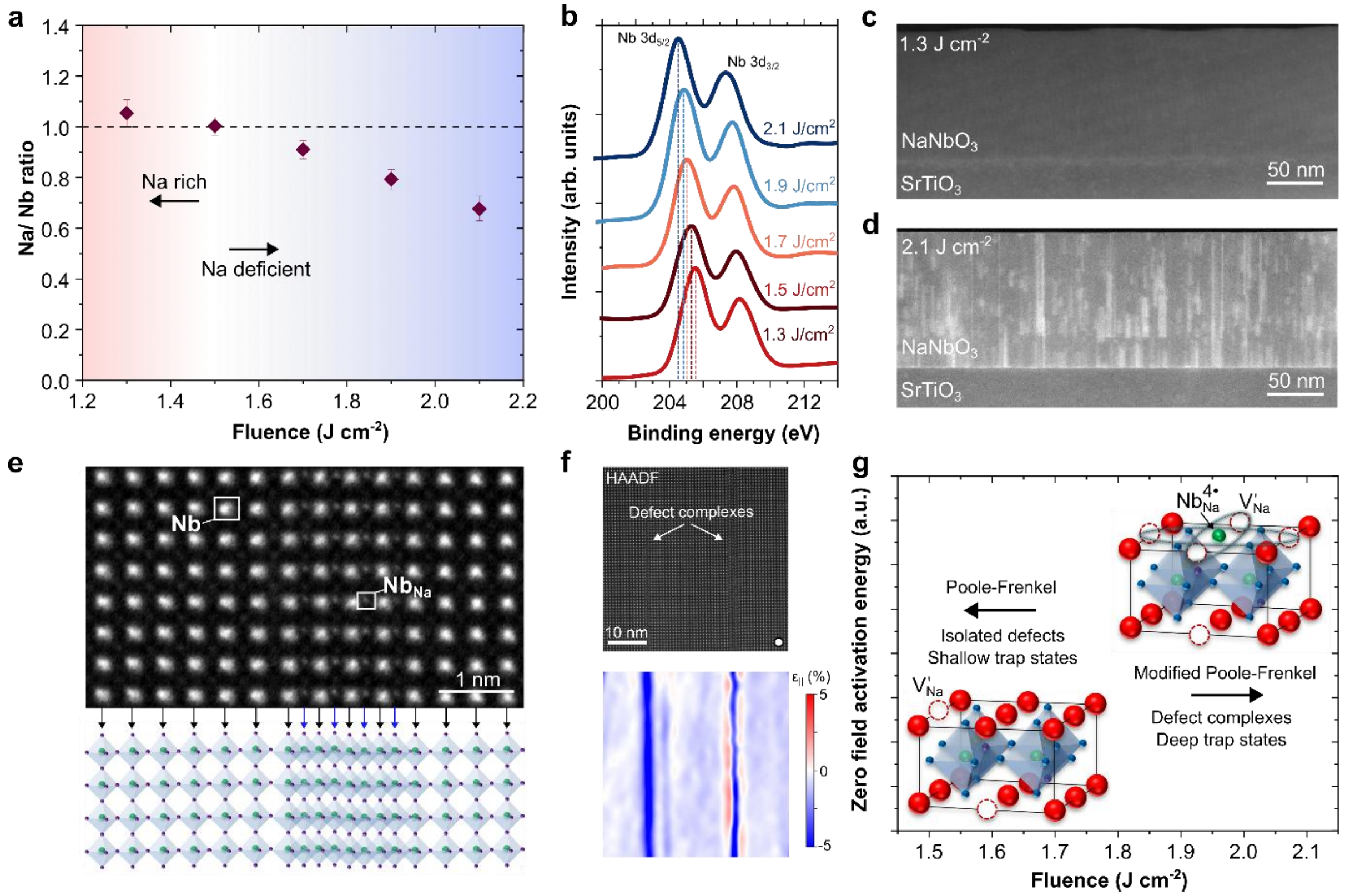

**Figure 3. Evolution of trap states in NaNbO3 films as a function of laser fluence. (a)** Na/Nb ratio as a function of laser fluence. **(b)** X-ray photoelectron spectroscopy (XPS) Nb 3*d* core-level spectra measured at different fluences, revealing a systematic peak shift toward lower binding energies with increasing fluence, indicative of Nb reduction. Cross-sectional low magnification LAADF-STEM images of $NaNbO_3$ films on $SrTiO_3$ substrates grown at **(c)** 1.3 and **(d)** 2.1 J cm$^{-2}$. **(e)** Cross-sectional high magnification HADDF-STEM image and the corresponding structural schematic for films grown at 2.1 J cm$^{-2}$. Extra columns associated with Nb antisite defects ($Nb_{Na}$) are highlighted by the blue arrows in the schematic. **(f)** The HAADF image and the corresponding longitudinal ($\varepsilon_{||}$) strain map. The real-space coarsening length for the strain map is denoted by the white circle in the HAADF image. **(g)** Schematic illustrating the evolution of trap states from isolated Na vacancies (shallow trap states) at low fluences to $Nb_{Na}^{\cdot\cdot\cdot\cdot} - 4V_{Na}'$ defect complexes (deep trap states) at high fluences.

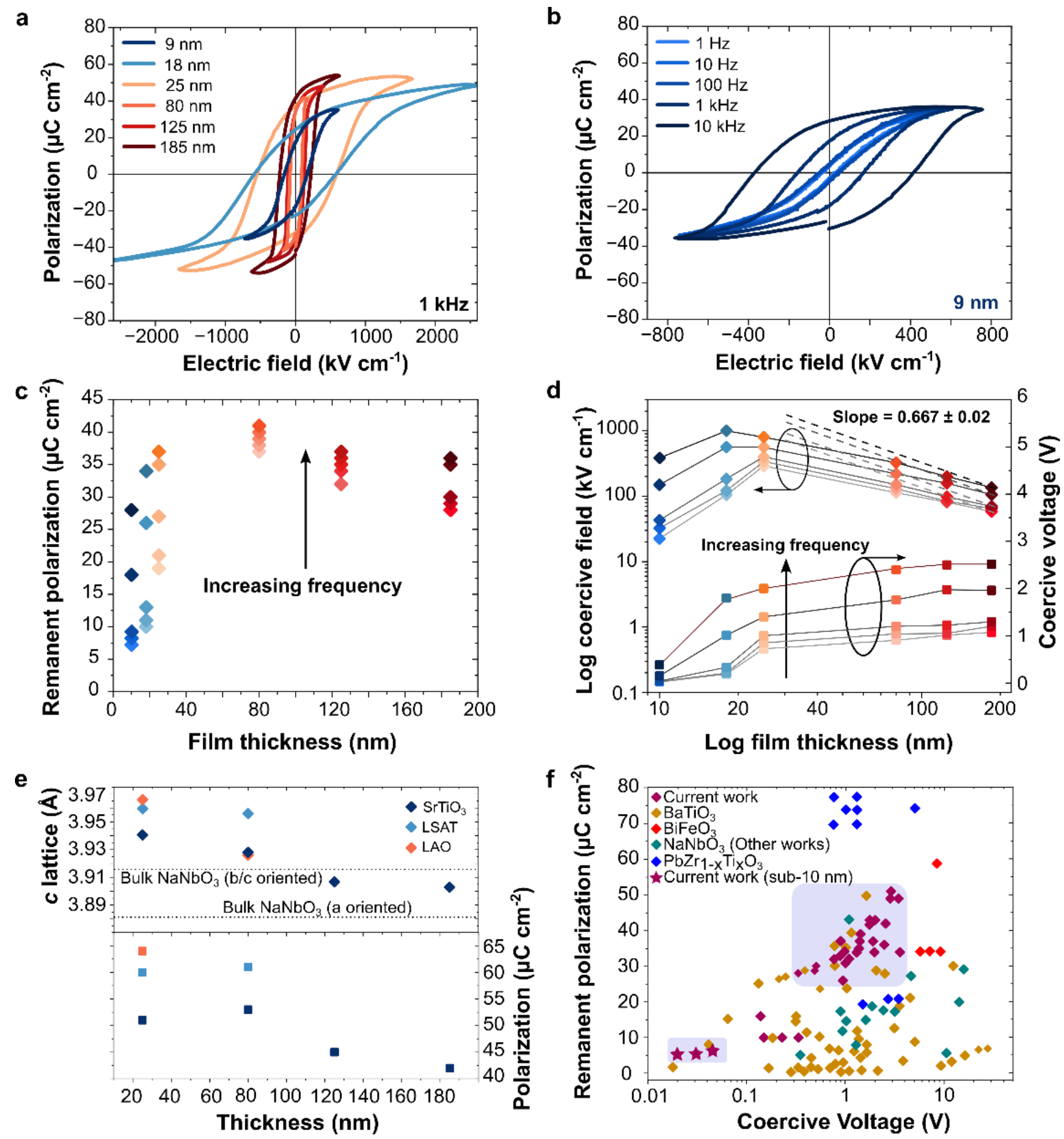

**Figure 4. Thickness scaling of high-fluence $NaNbO_3$ films. (a)** Thickness-dependent *P-E* hysteresis loops, measured at 1 kHz for films synthesized at 2.1 J $cm^{-2}$. **(b)** Frequency-dependent *P-E* hysteresis loops measured from 1 Hz to 10 kHz for 9 nm thick films. **(c)** Remanent polarization as a function of film thickness measured over frequencies ranging from 1 Hz to 10 kHz. **(d)** Thickness-dependent coercive fields (diamonds, left axis) and coercive voltages (squares, right axis) measured from 1 Hz to 10 kHz. Dashed lines represent linear fits to the traditional Janovec-Kay-Dunn (JKD) scaling law, revealing clear deviations for films thinner than 25 nm. **(e)** Out-of-plane *c*-lattice parameter (diamonds, top panel) and polarization (squares, bottom panel) of $NaNbO_3$ films grown on different substrates as a function of film thickness. Horizontal dotted lines indicate the out-of-plane lattice parameters of bulk $NaNbO_3$. **(f)** Comparison of remanent polarization versus coercive voltage for $NaNbO_3$ thin films in this work (highlighted in purple shade) and previously reported high-performance (001)-oriented perovskite ferroelectric thin films. Stars denote the sub-100 mV coercive voltages achieved in 9 nm thick $NaNbO_3$ films.